\documentstyle[prb,aps,floats,epsbox]{revtex}

\draft

\newcommand{\vect}[1]{{\mbox{\boldmath $#1$}}}

\newcommand{\beq}{\begin{equation}}
\newcommand{\eeq}{\end{equation}}
\newcommand{\beqa}{\begin{eqnarray}}
\newcommand{\eeqa}{\end{eqnarray}}

\newcommand{\Eq}[1]{(\ref{eq:#1})}

\newcommand{\Fig}[1]{Fig.~\ref{fg:#1}}

\newcommand{\Figure}[1]{Figure~\ref{fg:#1}}

\newcommand{\ct}[1]{\cite{#1}}

\newcommand{\eqref}[1]{(\ref{#1})}

\newcommand{\Deffig}[5]{
\begin{figure}[t]
\begin{center}
\null\ 
\postscriptbox{#4}{#3}{#2}
\end{center}
\caption[*]{#5}
\label{#1}
\end{figure}
}

\newcommand{\fgRawData}{
\Deffig{fg:RawData}{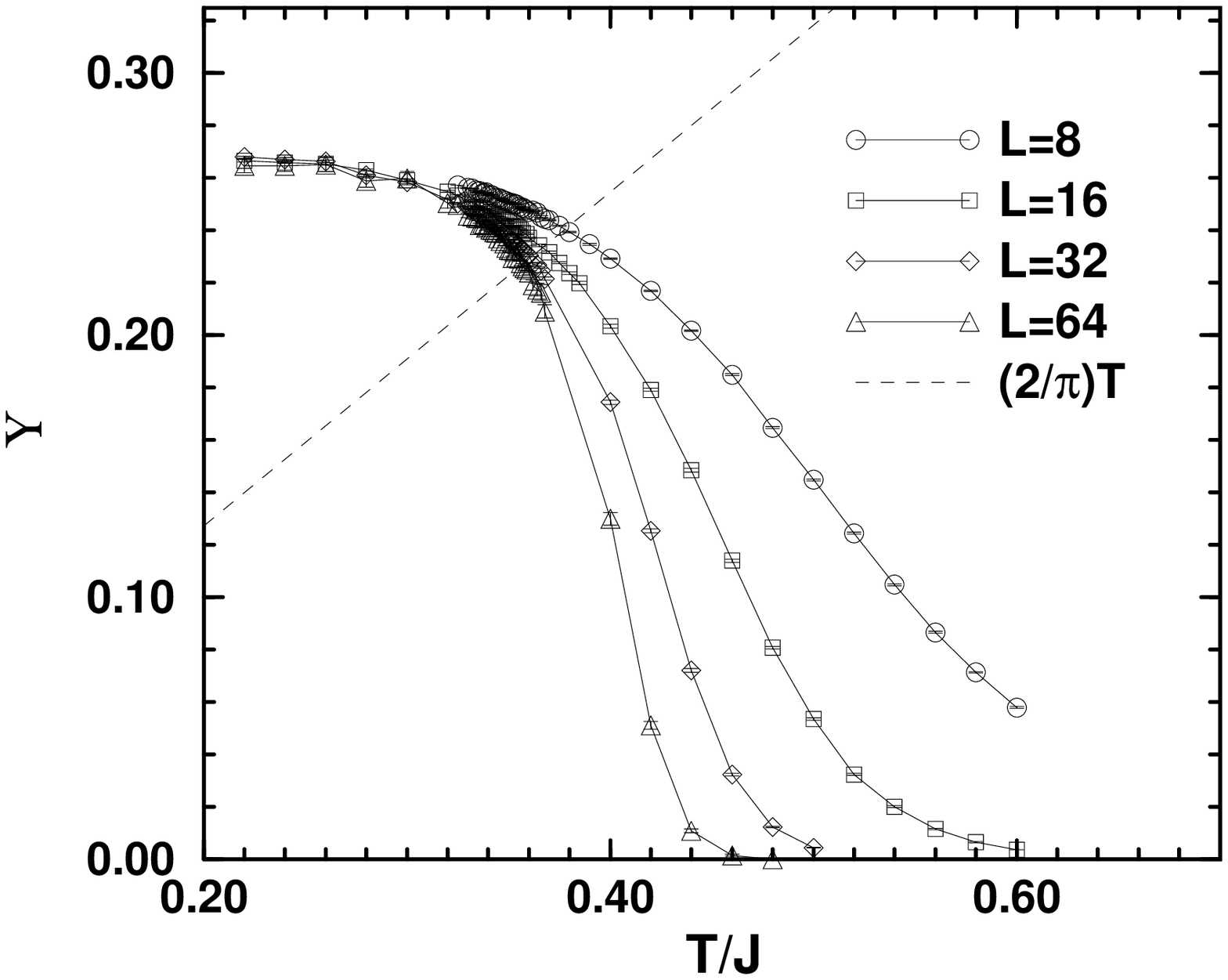}{63mm}{84mm}
{
The helicity modulus (or super fluid density)
$\Upsilon = (T/2) \langle {\bf W}^2 \rangle$
as a function of temperature.
The universal jump is expected at the point where $\Upsilon = 2T/\pi$.
Error bars are drown but most of them are so small that
they cannot be recognized.
} 
}

\newcommand{\fgScaling}{
\Deffig{fg:Scaling}{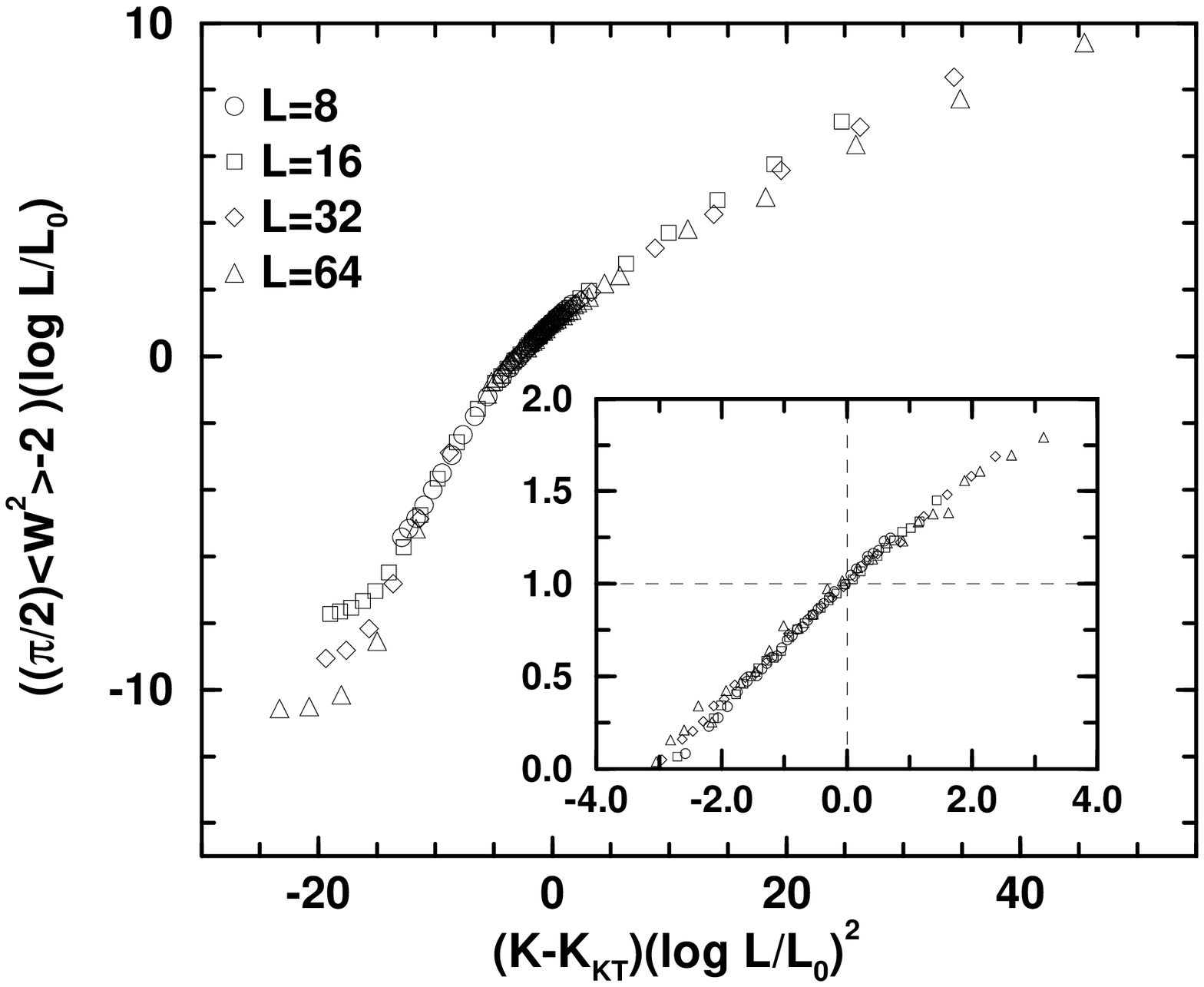}{63mm}{84mm}
{
A rescaled plot of the winding number fluctuation.
The inset is a enlarged view of the critical point.
} 
}

\newcommand{\fgA}{
\Deffig{fg:A}{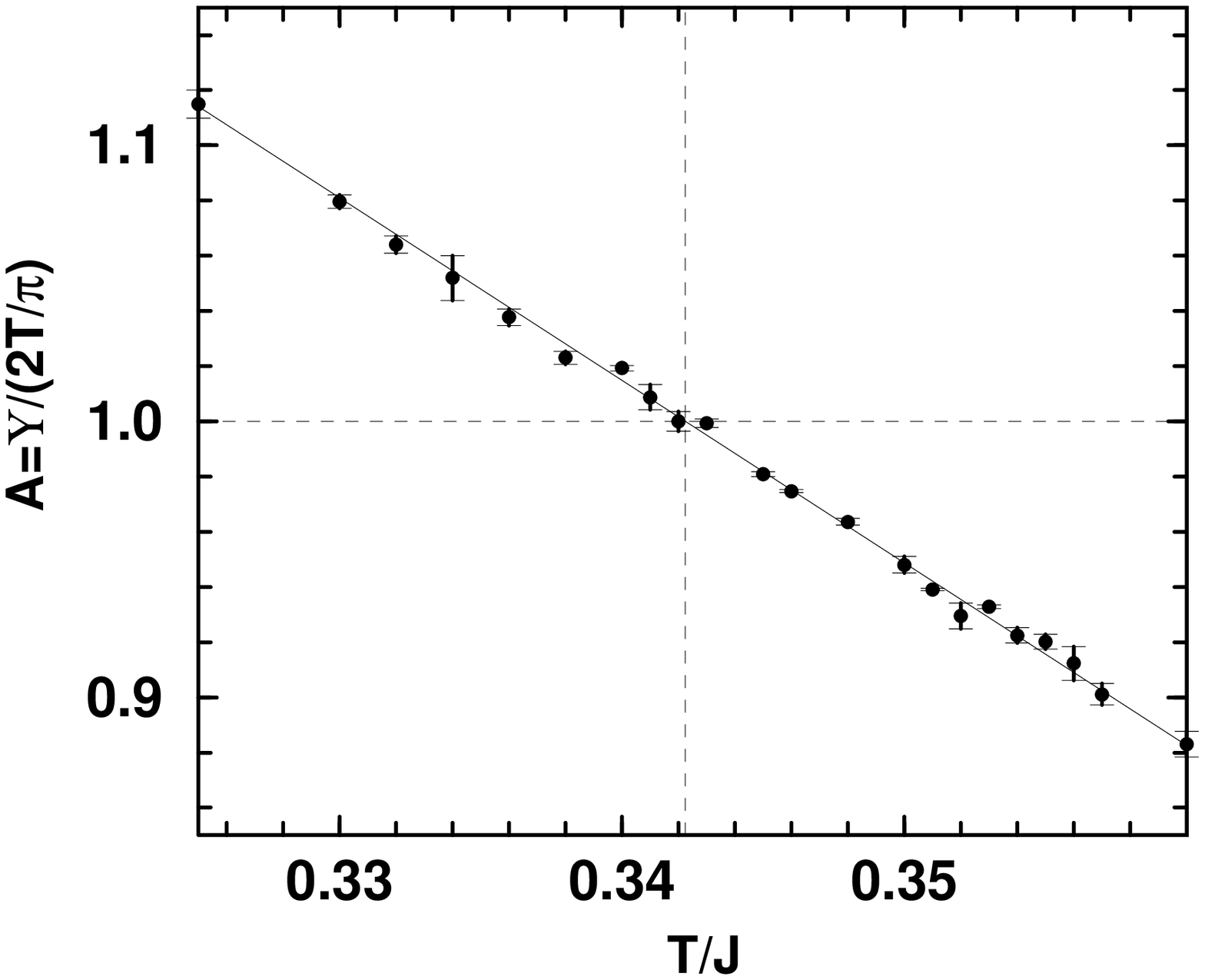}{60mm}{84mm}
{
The helicity modulus divided by the magnitude of its universal jump.
The dashed vertical line indicates the critical temperature at which
the solid line crosses the dashed horizontal line ($A=1$).
The solid line is determined by a linear fitting.
} 
}

\begin{document}

\twocolumn[\hsize\textwidth\columnwidth\hsize\csname@twocolumnfalse%
\endcsname
\title{
Universal Jump in the Helicity Modulus of the Two-Dimensional Quantum $XY$ Model
}

\author{ Kenji~Harada$^a$ and Naoki Kawashima$^b$ }
\address{$^a$ Division of Applied Systems Science, Kyoto University, Sakyo-ku,
    Kyoto 606-01, Japan}
\address{$^b$ Department of Physics, Toho University, Miyama 2-2-1, 
    Funabashi 274, Japan}

\date{11 Feburary 1997}

\maketitle

\begin{abstract}
The helicity modulus of the $S=1/2$ $XY$ model is precisely estimated
through a world line quantum Monte Carlo method
enhanced by a cluster update algorithm.
The obtained estimates for various system sizes and temperatures 
are well fitted by a scaling form with $L$ replaced by $\log (L/L_0)$,
which is inferred from the solution of the Kosterlitz 
renormalization group equation.
The validity of the Kosterlitz-Thouless theory for this model is confirmed.
\end{abstract}
%\pacs{PACS numbers: 05.70.Fh, 64.60.C, 75.40.Mg}
\vskip 0.5 cm
\vskip 0.5 cm
]

The nature of phase transition of the quantum $XY$ model in
two dimensions has not been fully clarified while the existence
of a transition at a finite temperature were suggested\ct{LohSG,Onogi}
some years ago.
Ding and Makivi\'c\ct{DingM} examined this problem 
systematically by a large scale Monte Carlo simulation
for the first time and concluded that a phase transition 
of the Kosterlitz-Thouless (KT)\ct{KT} type takes place at
$T_{\rm KT} = 0.350(4)$\ct{DingM} or $0.353(3)$\ct{Ding}.
In their calculation, the temperature dependence of the correlation 
length and the linear susceptibility were studied.
However, it is technically difficult to distinguish
an exponential divergence from an algebraic one.
Because of this difficulty, the validity of their conclusion on the
nature of the phase transition was questioned\ct{Comment}.
For the same reason, Gupta and Baillie\ct{GuptaB} did not conclude,
in spite of their very extensive Monte Carlo calculation,
that the phase transition of the classical $XY$ model is exactly
what the KT theory predicts.

In previous studies, to our knowledge,
a systematic study of the system size dependence of various
quantities was missing for the quantum $XY$ model.
On the other hand, 
for other models with a KT phase transition,
there are a number of reports on this size dependence.
For example, Sol\'yom and Ziman\ct{Solyom} studied size dependence of
the first excitation gap in the $S=1/2$ anisotropic $XXZ$ model
in one dimension, which is exactly solvable and known to have 
a transition of the KT type at the antiferromagnetic isotropic point.
They found that the exact estimates for the finite systems
do not fit into the standard form of the finite-size scaling
at the critical point.
Instead of using the ordinary finite-size scaling form,
Weber and Minnhagen\ct{WeberM} used the Kosterlitz
renormalization equation\ct{Kosterlitz} for the data analysis
in their study of the classical $XY$ model in two dimensions.
They verified the KT type phase transition by comparing
the size dependence of the helicity modulus, $\Upsilon$, with the solutions 
\underline{of} the renormalization group equation at the critical temperature.
They found a remarkable agreement extending even to the prefactor 
of the logarithmic correction.
Following the same idea, Olsson\ct{Olsson} performed a more detailed
analysis of the classical model
with a more extensive Monte Carlo simulation.
He found the resulting data including off-critical ones
consistent with the Kosterlitz renormalization group equation.

The helicity modulus is known to exhibit the universal jump
at the critical temperature\ct{NelsonK}.
This quantity corresponds to the super-fluid density when the model 
is regarded as a Boson system with hard-cores.
In the world-line quantum Monte Carlo method\ct{Suzuki},
the helicity modulus is related to the fluctuation in the 
total winding number of world-lines by the following equation\ct{Ceperley}.
\beq
  \Upsilon = (T/2) \langle \vect{W}^2 \rangle
\eeq
where $\vect{W} \equiv (W_x,W_y)$ with $W_x$ ($W_y$) being
the total winding number in the $x$ ($y$) direction.
Makivi\'c\ct{Makivic} computed this quantity
by means of a conventional world-line quantum Monte Carlo method.
The method is not ergodic\ct{LoopAlgorithm}, however, in that 
it does not allow the winding number to vary.
Therefore, Makivi\'c resorted to an alternative definition of the
quantity although it does not results in exactly the same
answer as the conventional definition does.
Another difficulty of the conventional Monte Carlo method is its 
long auto-correlation time near and below the critical temperature.
These difficulties limited the accuracy and the precision of 
the Makivi\'c's estimates and narrowed the temperature-range of the simulation.
To overcome these difficulties,
we use the cluster Monte Carlo method\ct{LoopAlgorithm} in the present note,
in which both the number of particles and the winding number can vary.
Another advantage of the method is its autocorrelation time which is
often shorter than that of the conventional method
by several orders of magnitude.

In the present note, we attempt at a detailed and precise comparison
between the model and the theory by Kosterlitz\ct{Kosterlitz} 
through an accurate estimation of the thermal fluctuation in the 
winding number near and below the critical temperature.
We will see that such an estimation allows us to
examine a new scaling form which is different from the ordinary 
\underline{finite-size scaling}. 
In this scaling form, the distance from the critical point, i.e.,
$K-K_{\rm KT}$ appears in the form of $(K-K_{\rm KT}) (\log (L/L_0))^2$,
in contrast to $(T-T_{\rm KT}) L^{y_T}$ in the 
ordinary finite-size scaling.
At the same time, the quantity 
$x \equiv \langle (\pi/2)\vect{W}^2 \rangle - 2$ scales 
as $x \log (L/L_0)$.

The $S=1/2$ quantum XY model is defined by the following Hamiltonian.
\begin{equation}
   H = - J \sum_{\langle ij \rangle} (S_i^x S_j^x + S_i^y S_j^y),
   \label{eq:Hamiltonian}
\end{equation}
where $\langle ij \rangle$ runs over all nearest-neighbor pairs on
the square lattice.
As for the spin operators, we here use the convention in which
$(S_i^{\mu})^2 = 1/4$ $(\mu = x,y,z)$.

In comparing the numerical result with the renormalization group theory,
the squared winding number is a useful quantity since it can be regarded as
the renormalized coupling constant and therefore its size dependence can be
directly predicted by the following renormalization equation \ct{Kosterlitz}.
\beq
  \frac{dx}{dl} = - y^2, \quad
  \frac{dy}{dl} = - xy \label{eq:RG}
\eeq
Here, $x$ and $y$ are renormalized parameters after a
renormalization operation up to the length scale $L\equiv L_0(T) e^{l}$ 
where $L_0(T)$ is some characteristic length of the order of the lattice 
constant and has no singularity at $T=T_{\rm KT}$.
The parameter $x$ is related to the renormalized coupling constant,
i.e., the helicity modulus by the following equation\ct{NelsonK}.
\beq
  x = \frac{\pi\Upsilon}{T} - 2 
    = \frac{\pi}{2}\langle \vect{W}^2 \rangle - 2.
\eeq
In what follows, we regard the estimate of
$(\pi/2)\langle \vect{W}^2 \rangle - 2$ for a system of 
the size $L$ at the temperature $T$ as $x(l=\log(L/L_0(T)),T)$.

The equation \Eq{RG} has an integral
\beq
  \Delta(T) \equiv x^2(l) - y^2(l) 
\eeq
which does not depend on $l \equiv \log (L/L_0(T))$.
As a function of $T$, this integral is non-singular even at $T=T_{\rm KT}$, i.e.,
$\Delta(K) = a (K-K_{\rm KT}) + b (K-K_{\rm KT})^2 + \cdots$ where
$K \equiv J/T$.
The solution of the renormalization equation \Eq{RG} is given by
\beq
  x(T,L) = \left\{
    \begin{array}{ll}
      \sqrt{|\Delta|} \coth ( \sqrt{|\Delta|} l ) & \quad (K > K_{\rm KT}) \\
      \sqrt{|\Delta|} \cot  ( \sqrt{|\Delta|} l ) & \quad (K < K_{\rm KT})
    \end{array} \right.\label{eq:Solution}
\eeq
Note that the ordinary finite size-scaling form
\beq
  x(T,L) = L^{\omega} g\left( \Delta L^{1/\nu} \right )
\eeq
cannot be consistent with \Eq{Solution}.
Instead, the solution \Eq{Solution} is a special case of the following 
form that one can obtain from the ordinary finite-size scaling 
form by replacing $L$ by $l = \log (L/L_0(T))$.
\beq
  x(T,L) = l^{-1} f(\Delta l^2). \label{eq:logFSS}
\eeq
When expressed in the form of \Eq{logFSS},
the solution \Eq{Solution} corresponds to a `scaling' function
$f(X)$ which is non-singular at $X=0$ in spite of the singularity
in $\sqrt{|\Delta|}$ at $T=T_{\rm KT}$.
To be more specific, the solution \Eq{Solution} corresponds to
\beq
  f(X) = 1 + O(X). \label{eq:TheCriticalValue}
\eeq
Note that the fitting functions used by Olsson\ct{Olsson}
(Equations (11a-c) with (16) and (18) in his paper) are consistent with this
scaling form.
Olsson's fitting function corresponds to
a temperature-independent $L_0(T)$ in our notation.

For our Monte Carlo simulation, using Suzuki-Trotter
decomposition, we transformed the two dimensional quantum XY model into
the (2+1)-dimensional Ising model with four spin plaquette interactions. 
The partition function is described as
\begin{equation}
  Z = \sum_{S} \prod_{p}\omega(S_{p}),
  \label{eq:6}
\end{equation}
where $S$ is the set of states $S_{i,t}^z$ on the (2+1)-dimensional lattice,
$S_{p}$ is a state of spins in a plaquette $p$ formed by four spins
$S_{i,t}^z,S_{j,t}^z,S_{i,t+1}^z,S_{j,t+1}^z$,
and $\omega(S_{p})$ is the two-spin propagator defined below. 
We apply periodic boundary conditions in all directions 
to preserve the translational invariance and 
to satisfy the trace requirements.

By numbering four local states on a bond as
$1 = (\uparrow\uparrow)$, $2 = (\uparrow \downarrow)$, 
$3 = (\downarrow \uparrow)$ and $4 = (\downarrow \downarrow)$,
the two-spin propagator can be written explicitly as
\begin{equation}
  \omega = \left(\begin{array}{cccc}
    1 & 0 & 0 & 0\\
    0 & \cosh ( \frac{\Delta\tau}{2} ) & \sinh ( \frac{\Delta\tau}{2} ) & 0\\
    0 & \sinh ( \frac{\Delta\tau}{2} ) & \cosh ( \frac{\Delta\tau}{2} ) & 0\\
    0 & 0 & 0 & 1
  \end{array}\right),
  \label{eq:5}
\end{equation}
where $\Delta\tau = J/(mk_{\rm B}T)$ with the Trotter number $m$,
and the temperature $T$.
In what follows, we will set $J=1$ and $k_{\rm B}=1$ for simplicity.

We have carried out a Monte Carlo simulation using the loop
algorithm\ct{LoopAlgorithm}.
In this algorithms, four spins in each plaquette are connected pair-wise.
The pairs are chosen stochastically.
The connected spins form loops which are units of flipping.
A world-line is a path connecting sites with up-spins which may 
wind around the system with periodic boundary condition. 
The winding numbers $W_x$ (or $W_y$) is defined as the sum of the numbers 
of such windings of all world-lines in the $x$ (or $y$) direction.
For example, $W_x$ can be rewritten as
\begin{equation}
  W_x = \frac{1}{L_x} \sum_p \alpha_x(S_{p})
  \label{eq:12}
\end{equation}
where $L_x$ is the lattice size in the $x$ direction.
The symbol $\alpha_x(S_{p})$ stands for the function which takes the 
value $1$ (or $-1$) if a world line passes through the plaquette $p$ 
in the positive (or negative) $x$-direction and the value 0, otherwise.
The other winding number $W_y$ is defined in the same manner.

In order to reduce the statistical error, we have used 
an improved estimator for the squared winding number of world lines.
It simply equals to one forth of the sum of squared winding numbers 
of loops formed by a clustering procedure (i.e., a graph assignment procedure)
in the cluster Monte Carlo method.
A more detailed discussion will be given elsewhere\ct{Future}.
In the present simulation, 
we have found that the new estimator reduces errors 
by about twenty percent.

In our simulations, we have taken various temperatures
between $0.22$ and $0.60$ and used lattices with $L=8,16,32$ and $64$.
As the Trotter number $m$, we have used $m=8,16$ and $32$.
When the systematic error due to the Trotter discretization exceeds
the statistical error, we have reduced the systematic error by the
extrapolation to $m=\infty$ using three data for different Trotter numbers.
The length of a typical run on $L=64$ at a temperature is
more than $2 \times 10^5$ Monte Carlo Sweeps(MCS). 
To make use of the vector processors,
we have developed the efficient vectorized code for the cluster identification.
The underlying idea for this vectorization was proposed by Mino\cite{Mino:91}
and it is based on the `divide-and-conquer' strategy.
In his algorithm, one firstly divides the lattice into small sub-lattices and,
at the same time, identifies clusters in each sub-lattices neglecting the
connectivity outside of the cluster.
One then deals with twice larger sub-lattices formed 
by a pair of previous sub-lattices.
By repeating this procedure, one can eventually get clusters identified
correctly in the whole lattice.
Using this algorithm, we have achieved the efficiency of
1.5 million site updates per second on the Fujitsu VPP500.

Each run is divided into several bins.
The length of a bin is taken so that bin averages are
statistically independent from each other at least approximately.
We have measured integrated autocorrelation times for $\vect{W}^2$ 
at low temperatures $T < 0.35J$ by the standard binning analysis
for a run of $10^5$ MCS with $L=64$. 
They turn out to be smaller than 2 MCS in all cases.
Since the smallest bin length used in the main calculation is $1500$ MCS,
the statistical independence among bins is assured. 
The statistical errors of various observables
have been calculated from the standard deviation of
the bin-average distribution.
\fgRawData
\fgScaling
\fgA

In \Fig{RawData}, the raw data for $\langle \vect{W}^2 \rangle \equiv 
\langle W_x^2 + W_y^2 \rangle$ are plotted.
\underline{\Figure{Scaling}} is its `scaling plot' using the the form of
\Eq{logFSS} assuming that $L_0(T)$ does not depend on the temperature,
which was a reasonably good approximation judging from the Olsson's
observation\ct{Olsson} (see Fig.14) for the classical model.
($L_0(T)$ in the present note corresponds to `const' in Eq.(18) in
Olsson's paper.)
The enlarged view around the critical point is shown as the inset.
The scaling plot \Fig{Scaling} is a reasonably good one,
considering the fact that we have only two adjustable parameters,
$K_{\rm KT}$ and $L_0$, while there are three such parameters in
the ordinary finite-size scaling plot.
The value of $K_{\rm KT}$ is determined so that the resulting plot
gives the `best' scaling plot.
To quantify the `goodness' of the scaling plot, 
a cost function $S(K_{\rm KT},L_0)$ is defined\ct{KawashimaI},
which is essentially the deviation from the locally linear fitting.
We used the data in the range
$0 < \langle \vect{W}^2 \rangle - 4/\pi < 0.25$,
which roughly corresponds to the temperature range $0.330 < T < 0.375$.
The best fitting is obtained with the transition temperature
\beq
  T_{\rm KT} = J/K_{\rm KT} = 0.342(2)J  \label{eq:TKT}
\eeq
which is significantly lower than the Ding and Makivi\'c's estimates
\ct{DingM,Ding}.
We confirmed that other choices of the range result in
the estimates of the critical temperature consistent with 
the value quoted above.
The best scaling plot is shown in \Fig{Scaling}.
It should be noted that the scaling function takes the value 1
at $K-K_{\rm KT}=0$ as is predicted by the Kosterlitz renormalization 
group equation (see \Eq{TheCriticalValue}).
Of course, an ordinary finite-size scaling does not predict
the value of the scaling function at this particular point.
Considering the fact that we have not used this information
while making the scaling plot,
the agreement between the scaling theory and the numerical result
is hardly understandable unless we assume the validity of the KT theory.
(Note that the criterion we have adopted in choosing the value of 
$K_{\rm KT}$ is simply that all points should fall onto a single curve.)
Therefore, we consider that the present result is a strong
confirmation of the validity of the Kosterlitz's scaling theory 
in the quantum $XY$ model in two dimensions.

We have also tried another analysis following Weber and Minnhagen
\ct{WeberM}.
At each temperature, we assumed the following system size dependence of
the helicity modulus.
\beq
  \frac{\pi\Upsilon}{2T} = \frac{\pi}{4} \langle \vect{W}^2 \rangle
  = A(T) \left( 1 + \frac1{2 \log (L/L_0(T))} \right)
  \label{eq:FittingFunction}
\eeq
This fitting form is correct at the critical point with $A(T_{\rm KT})=1$
(\Eq{logFSS} and \Eq{TheCriticalValue}).
Since the number of data at each temperature is small,
the critical temperature cannot be determined as the one
at which the fitting is best.
Instead, we plot the coefficient $A(T)$ as a function of temperature (\Fig{A}).
The critical temperature, then, is estimated as the point at which
$A(T)$ takes on the value 1.
Linear fitting of the data yields
\beq
  T_{\rm KT} = 0.3423(3)J,
\eeq
which is consistent with \Eq{TKT}.
(It is natural that here we have a more precise value than \Eq{TKT},
since we have assumed \Eq{FittingFunction}, a more specific fitting function
than \Eq{logFSS}.)

To summarize,
the loop algorithm have proven to be very efficient in studying
the winding number particularly at low temperatures, and
we have confirmed that the phase 
transition in the quantum $S=1/2$ $XY$ model is of the KT type
avoiding the subtle and technically difficult comparison between an
exponential divergence and an algebraic one.
We have demonstrated that not only the magnitude of the jump but also
very fine points in the Kosterlitz-Thouless-Nelson theory are realized in
the present model.

The authors would like to thank K.~Nomura for useful
comments and discussions.  The computations in the present
work were performed on Fujitsu's VPP500 at Kyoto University
Data Processing Center and partially at ISSP, the university
of Tokyo.

\end{document}